\documentclass[conference]{IEEEtran}
\ifCLASSINFOpdf
\else
\fi
\usepackage{url}

\usepackage[cmex10]{amsmath}
\usepackage{amssymb}

\usepackage{amsthm}

\theoremstyle{plain}
\newtheorem{theorem}{Theorem}

\newtheorem{definition}[theorem]{Definition}

\newtheorem{example}[theorem]{Example}

\newtheorem{replacements}{Replacements}

\newcommand{\abs}[1]{\left\lvert#1\right\rvert}

\DeclareMathOperator{\Dom}{dom}
\newcommand{\rest}[2]{#1\!\!\restriction_{#2}}

\newcommand{\N}{\mathbb{N}}
\newcommand{\Q}{\mathbb{Q}}
\newcommand{\R}{\mathbb{R}}
\newcommand{\X}{\{0,1\}^*}

\newcommand{\noi}{\noindent}

\begin{document}
%
\title{A statistical mechanical interpretation of
algorithmic information theory III:\\
Composite systems and fixed points}

\author{\IEEEauthorblockN{Kohtaro Tadaki}
\IEEEauthorblockA{Research and Development Initiative, Chuo University\\
JST CREST\\
1-13-27 Kasuga, Bunkyo-ku, Tokyo 112-8551, Japan\\
Email: tadaki@kc.chuo-u.ac.jp\quad
WWW: http://www2.odn.ne.jp/tadaki/}}


%


\maketitle

\begin{abstract}
\boldmath
The statistical mechanical interpretation of algorithmic information theory
(AIT, for short) was introduced and developed by our former works
[K.~Tadaki, Local Proceedings of CiE 2008, pp.~425--434, 2008]
and
[K.~Tadaki, Proceedings of LFCS'09, Springer's LNCS, vol.\ 5407,
pp.~422--440, 2009],
where we introduced the notion of thermodynamic quantities,
such as partition function $Z(T)$, free energy $F(T)$, energy $E(T)$,
and statistical mechanical entropy $S(T)$, into AIT.
We then discovered that, in the interpretation,
the temperature $T$ equals to the partial randomness of
the values of all these thermodynamic quantities,
where the notion of partial randomness is a stronger representation of
the compression rate by means of program-size complexity.
Furthermore,
we showed that this situation holds for the temperature itself
as a thermodynamic quantity,
namely,
for each of all the thermodynamic quantities above,
the computability of its value  at temperature $T$ gives
a sufficient condition for $T\in(0,1)$ to be a fixed point on partial randomness.
In this paper,
we develop the statistical mechanical interpretation of AIT further
and pursue its formal correspondence to normal statistical mechanics.
The thermodynamic quantities in AIT are defined
based on the halting set of an optimal computer,
which is a universal decoding algorithm used
to define the notion of program-size complexity.
We show that
there are infinitely many optimal computers which
give completely different sufficient conditions
in each of the thermodynamic quantities in AIT.
We do this by introducing the notion of composition of computers
into
AIT,
which corresponds to the notion of composition of systems
in normal statistical mechanics.
\end{abstract}


%
\IEEEpeerreviewmaketitle

\section{Introduction}

Algorithmic information theory (AIT, for short) is a framework
for applying
information-theoretic and probabilistic ideas to recursive function theory.
One of the primary concepts of AIT is the \textit{program-size complexity}
(or \textit{Kolmogorov complexity}) $H(s)$ of a finite binary string $s$,
which is defined as the length of the shortest binary
program
for an optimal computer
to output $s$.
Here an optimal computer is a universal decoding algorithm.
By the definition,
$H(s)$
is thought to represent
the amount of randomness contained in a finite binary string $s$,
which cannot be captured in a computational manner.
In particular,
the notion of program-size complexity plays a crucial role in
characterizing the \textit{randomness} of an infinite binary string,
or equivalently, a real.

In \cite{T08CiE}
we introduced and developed
a statistical mechanical interpretation of AIT.
We there introduced the notion of
\textit{thermodynamic quantities at temperature $T$},
such as partition function $Z(T)$, free energy $F(T)$, energy $E(T)$,
and statistical mechanical entropy $S(T)$,
into AIT.
These quantities are real functions of a real argument $T>0$.
We then proved that
if the temperature $T$ is a computable real with $0<T<1$
then, for each of these thermodynamic quantities,
the partial randomness of its value equals to $T$,
where the notion of \textit{partial randomness} is
a stronger representation of the compression rate
by means of program-size complexity.
Thus,
the temperature $T$ plays a role as
the partial randomness of all the thermodynamic quantities
in the statistical mechanical interpretation of AIT.
In \cite{T08CiE}
we further showed that
the temperature $T$ plays a role as the partial randomness of
the temperature $T$ itself,
which is a thermodynamic quantity of itself.
Namely,
we proved \textit{the fixed point theorem on partial randomness},%
\footnote{
The fixed point theorem on partial randomness is called
a fixed point theorem on compression rate in \cite{T08CiE}.}
which states that, for every $T\in(0,1)$,
if the value of partition function $Z(T)$ at temperature $T$
is a computable real,
then the partial randomness of $T$ equals to $T$,
and therefore the compression rate of $T$ equals to $T$,
i.e.,
$\lim_{n\to\infty}H(\rest{T}{n})/n=T$,
where $\rest{T}{n}$ is the first $n$ bits of the base-two expansion of $T$.

In our second work \cite{T09LFCS} on this
interpretation,
we showed that
a fixed point theorem
of the same form
as for $Z(T)$
holds also for each of free energy $F(T)$, energy $E(T)$, and
statistical mechanical entropy $S(T)$.
Moreover,
based on
the statistical mechanical relation $F(T)=-T\log_2 Z(T)$,
we showed that the computability of $F(T)$ gives
completely different fixed points from the computability of $Z(T)$.

In this paper,
we develop the statistical mechanical interpretation of AIT further
and pursue its formal correspondence to normal statistical mechanics.
As a result,
we unlock the properties of the sufficient conditions further.
The thermodynamic quantities in AIT are defined
based on the halting set of an optimal computer.
In this paper,
we show in Theorem~\ref{main} below that
there are infinitely many optimal computers which
give completely different sufficient conditions
in each of the thermodynamic quantities in AIT.
We do this by introducing the notion of composition of computers
into
AIT,
which corresponds to the notion of composition of systems
in normal statistical mechanics.

\section{Preliminaries}
\label{preliminaries}

\subsection{Basic notation}
\label{basic notation}

We start with some notation about numbers and strings
which will be used in this paper.
$\N=\left\{0,1,2,3,\dotsc\right\}$ is the set of natural numbers,
and $\N^+$ is the set of positive integers.
$\Q$ is the set of rationals, and
$\R$ is the set of reals.
Let $f\colon S\to\R$ with $S\subset\R$.
We say that $f$ is \textit{increasing} (resp., \textit{decreasing})
if $f(x)<f(y)$ (resp., $f(x)>f(y)$) for all $x,y\in S$ with $x<y$.

Normally, $o(n)$ denotes any
function $f\colon \N^+\to\R$ such
that $\lim_{n \to \infty}f(n)/n=0$.

$\X=
\left\{
  \lambda,0,1,00,01,10,11,000,\dotsc
\right\}$
is the set of finite binary strings,
where $\lambda$ denotes the \textit{empty string}.
For any $s \in \X$, $\abs{s}$ is the \textit{length} of $s$.
A subset $S$ of $\X$ is called
\textit{prefix-free}
if no string in $S$ is a prefix of another string in $S$.
For any partial function $f$,
the domain of definition of $f$ is denoted by $\Dom f$.

Let $\alpha$ be an arbitrary real.
We denote by $\rest{\alpha}{n}\in\X$
the first $n$ bits of the base-two expansion of
$\alpha - \lfloor \alpha \rfloor$ with infinitely many zeros,
where $\lfloor \alpha \rfloor$ is the greatest integer
less than or equal to $\alpha$.
For example,
in the case of $\alpha=5/8$,
$\rest{\alpha}{6}=101000$.

We say that a real $\alpha$ is \textit{computable} if
there exists a total recursive function $f\colon\N^+\to\Q$ such that
$\abs{\alpha-f(n)} < 1/n$ for all $n\in\N^+$.
See e.g.~Weihrauch \cite{W00}
for the detail of the treatment of
the computability of reals.

\subsection{Algorithmic information theory}
\label{ait}

In the following
we concisely review some definitions and results of
algorithmic information theory
\cite{C75,C87b,N09,DH09}.
A \textit{computer} is a partial recursive function
$C\colon \X\to \X$
such that
$\Dom C$ is a
nonempty
prefix-free set.
For each computer $C$ and each $s \in \X$,
$H_C(s)$ is defined by
$H_C(s) =
\min
\left\{\,
  \abs{p}\,\big|\;p \in \X\>\&\>C(p)=s
\,\right\}$
(may be $\infty$).
A computer $U$ is said to be \textit{optimal} if
for each computer $C$ there exists $d\in\N$,
which depends on $C$,
with the following property;
for every $p\in\Dom C$ there exists $q\in\X$ for which
$U(q)=C(p)$ and $\abs{q}\le\abs{p}+d$.
It is easy to see that there exists an optimal computer.
We choose a particular optimal computer $U$
as the standard one for use,
and define $H(s)$ as $H_U(s)$,
which is referred to as
the \textit{program-size complexity} of $s$ or
the \textit{Kolmogorov complexity} of $s$.
It follows that
for every computer $C$ there exists $d\in\N$ such that,
for every $s\in\X$,
$H(s) \le H_C(s)+d$.

For any $\alpha\in\R$,
we say that $\alpha$ is \textit{weakly Chaitin random}
if there exists $c\in\N$ such that
$n-c\le H(\rest{\alpha}{n})$ for all $n\in\N^+$
\cite{C75,C87b}.
On the other hand,
for any $\alpha\in\R$,
we say that $\alpha$ is
\textit{Chaitin random}
if $\lim_{n\to \infty} H(\rest{\alpha}{n})-n=\infty$ \cite{C75,C87b}.
Obviously,
for every $\alpha\in\R$,
if $\alpha$ is Chaitin random,
then $\alpha$ is weakly Chaitin random.
We can show that
the converse also hold.
Thus,
for every $\alpha\in\R$,
$\alpha$ is weakly Chaitin random if and only if
$\alpha$ is Chaitin random
(see
Chaitin
\cite{C87b} for the proof and historical detail).

\vspace*{-2mm}

\subsection{Partial randomness}
\label{partial}

In the works \cite{T99,T02},
we generalized the notion of
the randomness of a real
so that \textit{the degree of the randomness},
which is often referred to as
\textit{the partial randomness} recently
\cite{CST06,RS05,CS06},
can be characterized by a real $T$
with $0\le T\le 1$ as follows.

\begin{definition}[weak Chaitin $T$-randomness]
Let
$T\in\R$ with
$T\ge 0$.
For any $\alpha\in\R$,
we say that $\alpha$ is \textit{weakly Chaitin $T$-random} if
there exists $c\in\N$ such that
$Tn-c \le H(\rest{\alpha}{n})$
for all $n\in\N^+$.
\hfill\IEEEQED
\end{definition}

\begin{definition}[$T$-compressibility]
Let
$T\in\R$ with
$T\ge 0$.
For any $\alpha\in\R$,
we say that $\alpha$ is \textit{$T$-compressible} if
$H(\rest{\alpha}{n})\le Tn+o(n)$,
which is equivalent to
$\limsup_{n \to \infty}H(\rest{\alpha}{n})/n\le T$.
\hfill\IEEEQED
\end{definition}

In the case of $T=1$,
the weak Chaitin $T$-randomness results in the weak Chaitin randomness.
For every $T\in[0,1]$ and every $\alpha\in\R$,
if $\alpha$ is weakly Chaitin $T$-random and $T$-compressible,
then
\vspace*{-2mm}
\begin{equation}\label{compression-rate}
  \lim_{n\to \infty} \frac{H(\rest{\alpha}{n})}{n} = T.
\end{equation}
The
left-hand side of \eqref{compression-rate}
is referred to as the \textit{compression rate} of
a real $\alpha$ in general.
Note, however, that \eqref{compression-rate}
does not necessarily imply that $\alpha$ is weakly Chaitin $T$-random.
Thus, the notion of partial randomness is
a stronger representation of
the notion of
compression rate.

\begin{definition}[Chaitin $T$-randomness, Tadaki \cite{T99,T02}]
  Let $T\in\R$ with $T\ge 0$.
  For any $\alpha \in\R$,
  we say that $\alpha$ is \textit{Chaitin $T$-random} if
  $\lim_{n\to \infty} H(\rest{\alpha}{n})-Tn = \infty$.
\hfill\IEEEQED
\end{definition}

In the case of $T=1$,
the Chaitin $T$-randomness results in the Chaitin randomness.
Obviously,
for every $T\in[0,1]$ and every $\alpha\in\R$,
if $\alpha$ is Chaitin $T$-random,
then $\alpha$ is weakly Chaitin $T$-random.
However,
in 2005 Reimann and Stephan \cite{RS05} showed that,
in the case of $T<1$,
the converse does not necessarily hold.
This contrasts with the
equivalence
between
the weak Chaitin randomness and the Chaitin randomness,
each of which corresponds to the case of $T=1$.

\section{The previous results}
\label{tcr}

In this section,
we review some results of the statistical mechanical interpretation of
AIT,
developed by our former works \cite{T08CiE,T09LFCS}.
We first introduce the notion of thermodynamic quantities into AIT
in the following manner.

In statistical mechanics,
the partition function $Z_{\mathrm{sm}}(T)$,
free energy $F_{\mathrm{sm}}(T)$,
energy $E_{\mathrm{sm}}(T)$,
and
entropy $S_{\mathrm{sm}}(T)$
at temperature $T$
are given
as follows:
\begin{equation}\label{tdqsm}
\begin{split}
  Z_{\mathrm{sm}}(T)
  &=\sum_{x\in X}e^{-\frac{E_x}{k_{\mathrm{B}}T}}, \\
  F_{\mathrm{sm}}(T)
  &=-k_{\mathrm{B}}T\ln Z_{\mathrm{sm}}(T), \\
  E_{\mathrm{sm}}(T)
  &=\frac{1}{Z_{\mathrm{sm}}(T)}\sum_{x\in X}E_xe^{-\frac{E_x}{k_{\mathrm{B}}T}}, \\
  S_{\mathrm{sm}}(T)
  &=\frac{E_{\mathrm{sm}}(T)-F_{\mathrm{sm}}(T)}{T},
\end{split}
\end{equation}
where $X$ is a complete set of energy eigenstates of
a quantum system
and $E_x$ is the energy of an energy eigenstate $x$.
The constant $k_{\mathrm{B}}$ is called the Boltzmann Constant,
and the $\ln$ denotes the natural logarithm.%
\footnote{
For the thermodynamic quantities in statistical mechanics,
see e.g.~Chapter 16 of \cite{C85}
and Chapter 2 of \cite{TKS92}.
To be precise,
the partition function is not a thermodynamic quantity
but a statistical mechanical quantity.
}

Let $C$ be an arbitrary computer.
We introduce the notion of thermodynamic quantities into AIT
by performing Replacements~\ref{CS06} below
for the thermodynamic quantities \eqref{tdqsm} in statistical mechanics.

\begin{replacements}\label{CS06}\hfill
\begin{enumerate}
  \item Replace the complete set $X$ of energy eigenstates $x$
    by the set $\Dom C$ of all programs $p$ for $C$.
  \item Replace the energy $E_x$ of an energy eigenstate $x$
    by the length $\abs{p}$ of a program $p$.
  \item Set the Boltzmann Constant $k_{\mathrm{B}}$ to $1/\ln 2$.
\hfill\IEEEQED
\end{enumerate}
\end{replacements}

Thus,
motivated by the formulae \eqref{tdqsm}
and taking into account Replacements~\ref{CS06},
we introduce the notion of thermodynamic quantities into AIT
as follows.

\begin{definition}[thermodynamic quantities in AIT,
\cite{T08CiE}]\label{tdqait}
Let $C$ be any computer,
and let $T$ be any real with $T>0$.

First consider the case where $\Dom C$ is an infinite set.
In this case,
we choose a particular enumeration
$p_1,p_2,p_3,p_4,\dotsc$ of the countably infinite set $\Dom C$.%
\footnote{
The enumeration $\{p_i\}$ can be chosen quite arbitrarily,
and the results of this paper are independent of the choice of $\{p_i\}$.
This is because
the sum $\sum_{i=1}^k 2^{-\abs{p_i}/T}$ and
$\sum_{i=1}^k \abs{p_i}2^{-\abs{p_i}/T}$
in Definition~\ref{tdqait}
are positive term series
and converge as $k\to\infty$ for every $T\in(0,1)$.
}
\begin{enumerate}
\item
  The \textit{partition function} $Z_C(T)$
  at temperature $T$
  is defined
  as $\lim_{k\to\infty} Z_k(T)$
  where
\vspace*{-1mm}
  \begin{equation}\label{DZkT}
    Z_k(T)=\sum_{i=1}^k 2^{-\frac{\abs{p_i}}{T}}.
  \end{equation}
\item
  The \textit{free energy} $F_C(T)$
  at temperature $T$
  is defined
  as $\lim_{k\to\infty} F_k(T)$
  where
\vspace*{-1mm}
  \begin{equation}\label{DFkT}
    F_k(T)=-T\log_2 Z_k(T).
  \end{equation}
\item
  The \textit{energy} $E_C(T)$
  at temperature $T$
  is defined
  as $\lim_{k\to\infty} E_k(T)$
  where
\vspace*{-1mm}
  \begin{equation}\label{DEkT}
    E_k(T)
    =\frac{1}{Z_k(T)}\sum_{i=1}^k \abs{p_i}2^{-\frac{\abs{p_i}}{T}}.
  \end{equation}
\item
  The \textit{statistical mechanical entropy} $S_C(T)$
  at temperature $T$
  is defined
  as $\lim_{k\to\infty} S_k(T)$
  where
  \begin{equation}\label{DSkT}
    S_k(T)=\frac{E_k(T)-F_k(T)}{T}.
  \end{equation}
\end{enumerate}

In the case where $\Dom C$ is a nonempty finite set,
the quantities
$Z_C(T)$, $F_C(T)$, $E_C(T)$, and $S_C(T)$ are just
defined as \eqref{DZkT}, \eqref{DFkT}, \eqref{DEkT}, and \eqref{DSkT},
respectively,
where $p_1,\dots,p_k$ is an enumeration of the finite set $\Dom C$.
\hfill\IEEEQED
\end{definition}

Note that,
for every optimal computer $V$,
$Z_V(1)$ is precisely a Chaitin $\Omega$ number
introduced by Chaitin \cite{C75}.
Theorems~\ref{cprpffe} and \ref{cpreesh} below
hold for these thermodynamic quantities in AIT.

\begin{theorem}[properties of $Z(T)$ and $F(T)$,
\cite{T99,T02,T08CiE}]\label{cprpffe}
Let $V$ be an optimal computer, and let $T\in\R$.
\begin{enumerate}
  \item If $0<T\le 1$ and $T$ is computable,
    then each of $Z_V(T)$ and $F_V(T)$ converges
    and is
    weakly Chaitin $T$-random and $T$-compressible.
  \item If $1<T$,
    then $Z_V(T)$ and $F_V(T)$ diverge to $\infty$ and $-\infty$,
    respectively.
\hfill\IEEEQED
\end{enumerate}
\end{theorem}

\begin{theorem}[properties of
$E(T)$ and $S(T)$,
\cite{T08CiE}]\label{cpreesh}
Let $V$ be an optimal computer, and let $T\in\R$.
\begin{enumerate}
  \item If $0<T<1$ and $T$ is computable,
    then each of
    $E_V(T)$ and $S_V(T)$
    converges
    and is
    Chaitin $T$-random and $T$-compressible.
  \item If $1\le T$,
    then both $E_V(T)$ and $S_V(T)$ diverge to $\infty$.
\hfill\IEEEQED
\end{enumerate}
\end{theorem}

The above
two
theorems show that
if $T$ is
a computable real with
$T\in(0,1)$
then the temperature $T$ equals to the partial randomness
(and therefore the compression rate) of
the values of all the thermodynamic quantities
in Definition~\ref{tdqait}
for an optimal computer.

These theorems also show that
the values of all the thermodynamic quantities
diverge when the temperature $T$ exceeds $1$.
This phenomenon might be regarded as
some sort of phase transition
in statistical mechanics.
Note here that
the weak Chaitin $T$-randomness in Theorem~\ref{cprpffe}
is replaced by the Chaitin $T$-randomness in Theorem~\ref{cpreesh}
in exchange for the divergence at $T=1$.

In statistical mechanics or thermodynamics,
among all thermodynamic quantities
one of the most typical thermodynamic quantities is temperature
itself.
Theorem~\ref{fptpr} below shows that
the partial randomness of the temperature $T$
can equal to the temperature $T$ itself
in the statistical mechanical interpretation of AIT.

We denote by $\mathcal{FP}_w$ the set of all real $T\in(0,1)$
such that $T$ is weakly Chaitin $T$-random and $T$-compressible,
and denote by $\mathcal{FP}$ the set of all real $T\in(0,1)$
such that $T$ is Chaitin $T$-random and $T$-compressible.
Obviously, $\mathcal{FP}\subset\mathcal{FP}_w$.
Each element $T$ of $\mathcal{FP}_w$ is
a \textit{fixed point on partial randomness},
i.e.,
satisfies the property that
the partial randomness of $T$ equals to $T$ itself,
and therefore satisfies that
$\lim_{n\to\infty}H(\rest{T}{n})/n=T$.
Let $V$ be a computer.
We define the sets $\mathcal{Z}(V)$ by
\begin{equation*}
  \mathcal{Z}(V)=\{\,T\in(0,1)\mid Z_V(T)\text{ is computable}\,\}.
\end{equation*}
In the same manner,
we define the sets $\mathcal{F}(V)$, $\mathcal{E}(V)$,
and $\mathcal{S}(V)$
based on the computability of
$F_V(T)$, $E_V(T)$, and $S_V(T)$,
respectively.
We can then
show the following.

\begin{theorem}[fixed points on partial randomness,
\cite{T08CiE,T09LFCS}]\label{fptpr}
Let $V$ be an optimal computer.
Then $\mathcal{Z}(V)\cup\mathcal{F}(V)\subset\mathcal{FP}_w$
and 
$\mathcal{E}(V)\cup\mathcal{S}(V)\subset\mathcal{FP}$.
\hfill\IEEEQED
\end{theorem}

Theorem~\ref{fptpr} is just a fixed point theorem on partial randomness,
where the computability of each of
the values $Z_V(T)$, $F_V(T)$, $E_V(T)$, and $S_V(T)$
gives a sufficient condition
for a real $T\in(0,1)$ to be a fixed point on partial randomness.
Thus,
by Theorem~\ref{fptpr},
the above observation
that the temperature $T$ equals to the partial randomness of
the values of the thermodynamic quantities
in the statistical mechanical interpretation of AIT
is further confirmed.

\section{The main result}
\label{imscs}

In this paper,
we investigate the properties of the sufficient conditions
for $T$ to be a fixed point on partial randomness
in Theorem~\ref{fptpr}.
Using the monotonicity and continuity of
the functions $Z_V(T)$ and $F_V(T)$ on temperature $T$
and using the statistical mechanical relation $F_V(T)=-T\log_2 Z_V(T)$,
which holds from Definition~\ref{tdqait},
we can show the following theorem for the sufficient conditions
in Theorem~\ref{fptpr}.

\begin{theorem}[\cite{T09LFCS}]\label{scZF}
Let $V$ be an optimal computer.
Then
each of the sets $\mathcal{Z}(V)$ and $\mathcal{F}(V)$
is dense in $(0,1)$
while $\mathcal{Z}(V)\cap\mathcal{F}(V)=\emptyset$.
\hfill\IEEEQED
\end{theorem}

Thus, for every optimal computer $V$,
the computability of $F_V(T)$ gives
completely different fixed points from the computability of $Z_V(T)$.
This implies also that
$\mathcal{Z}(V)\subsetneqq\mathcal{FP}_w$
and $\mathcal{F}(V)\subsetneqq\mathcal{FP}_w$.

The aim of this paper is
to investigate the structure of $\mathcal{FP}_w$ and $\mathcal{FP}$
in greater detail.
Namely,
we show in Theorem~\ref{main} below that
there are infinitely many optimal computers which
give completely different sufficient conditions
in each of the thermodynamic quantities in AIT.
We say that an infinite sequence $V_1,V_2,V_3,\dotsc$ of computers
is \textit{recursive} if
there exists a partial recursive function $F\colon\N^+\times\X\to\X$
such that for each $n\in\N^+$ the following two hold:
(i) $p\in\Dom V_n$ if and only if $(n,p)\in\Dom F$, and
(ii) $V_n(p)=F(n,p)$ for every $p\in\Dom V_n$.
Then the main result of this paper is given as follows.

\begin{theorem}[main result]\label{main}
There exists a recursive infinite sequence $V_1,V_2,V_3,\dotsc$ of
optimal computers which satisfies the following conditions:
\begin{enumerate}
\item $\mathcal{Z}(V_i)\cap\mathcal{Z}(V_j)=
  \mathcal{F}(V_i)\cap\mathcal{F}(V_j)=
  \mathcal{E}(V_i)\cap\mathcal{E}(V_j)=
  \mathcal{S}(V_i)\cap\mathcal{S}(V_j)=\emptyset$
for all $i,j$ with $i\neq j$.
\item $\bigcup_{i}\mathcal{Z}(V_i)\subset\mathcal{FP}_w$
  and $\bigcup_{i}\mathcal{F}(V_i)\subset\mathcal{FP}_w$.
\item $\bigcup_{i}\mathcal{E}(V_i)\subset\mathcal{FP}$
  and $\bigcup_{i}\mathcal{S}(V_i)\subset\mathcal{FP}$.
\hfill\IEEEQED
\end{enumerate}
\end{theorem}

In the subsequent sections we prove the above theorems
by introducing the notion of \textit{composition} of computers
into
AIT,
which corresponds to the notion of composition of systems
in normal statistical mechanics.

\section{Composition of computers}
\label{compsite}

\begin{definition}[composition of computers]\hfill\\
Let $C_1,C_2,\dots,C_N$ be computers.
The composition $C_1\oslash C_2\oslash \dotsm \oslash C_N$
of $C_1$, $C_2$, $\dotsc$, and $C_N$
is defined as the computer $D$ such that
(i) $\Dom D=\{p_1p_2\dots p_N \mid
p_1\in\Dom C_1\ \&\ p_2\in\Dom C_2\ \&\ \dotsm\ \&\ p_N\in\Dom C_N\}$,
and (ii) $D(p_1p_2\dots p_N)=C_1(p_1)$
for every $p_1\in\Dom C_1$, $p_2\in\Dom C_2$, $\dotsc$, and $p_N\in\Dom C_N$.
\hfill\IEEEQED
\end{definition}

\begin{theorem}\label{ocmpo}
Let $C_1,C_2,\dots,C_N$ be computers.
If $C_1$ is optimal
then $C_1\oslash C_2\oslash \dotsm \oslash C_N$ is also optimal.
\end{theorem}

\begin{proof}
We first choose particular strings $r_2$, $r_3$, $\dotsc$, $r_N$ with
$r_2\in\Dom C_2$, $r_3\in\Dom C_3$, $\dotsc$, and $r_N\in\Dom C_N$.
Let $C$ be an arbitrary computer.
Then, by the definition of the optimality of $C_1$,
there exists $d\in\N$ with the following property;
for every $p\in\Dom C$ there exists $q\in\X$ for which
$C_1(q)=C(p)$ and $\abs{q}\le\abs{p}+d$.
It follows from the definition of the composition
$C_1\oslash C_2\oslash \dotsm \oslash C_N$
that for every $p\in\Dom C$ there exists $q\in\X$ for which
$(C_1\oslash C_2\oslash \dotsm \oslash C_N)(qr_2r_3\dots r_N)=C(p)$
and
$\abs{qr_2r_3\dots r_N}\le \abs{p}+\abs{r_2r_3\dots r_N}+d$.
Thus $C_1\oslash C_2\oslash \dotsm \oslash C_N$ is an optimal computer.
\end{proof}

In the same manner as in normal statistical mechanics,
we can prove Theorem~\ref{quc} below
for the thermodynamic quantities in AIT.
In particular,
the equations~\eqref{cmpsfe}, \eqref{cmpse}, and \eqref{cmpss}
correspond to the fact that
free energy, energy, and entropy are extensive parameters
in thermodynamics, respectively. 

\begin{theorem}\label{quc}
Let $C_1,C_2,\dots,C_N$ be computers.
Then the following hold for every $T\in(0,1)$.
\vspace*{-1mm}
\begin{align}
  Z_{C_1\oslash \dotsm \oslash C_N}(T)&=Z_{C_1}(T)\dotsm Z_{C_N}(T),
  \nonumber\\
  F_{C_1\oslash \dotsm \oslash C_N}(T)&=F_{C_1}(T)+\dotsm +F_{C_N}(T),
  \label{cmpsfe}\\
  E_{C_1\oslash \dotsm \oslash C_N}(T)&=E_{C_1}(T)+\dotsm +E_{C_N}(T),
  \label{cmpse}\\
  S_{C_1\oslash \dotsm \oslash C_N}(T)&=S_{C_1}(T)+\dotsm +S_{C_N}(T).
  \label{cmpss}
\end{align}
\hfill\IEEEQED
\end{theorem}

For any computer $C$ and any $n\in\N^+$,
the computer $\underbrace{C\oslash\dotsm\oslash C}_{n}$
is denoted by $C^{\oslash n}$.

\section{The proof of the main result}
\label{proof-mrs}

In order to prove the main result, Theorem~\ref{main},
we
also
introduce the notion of \textit{physically reasonable computer}.

\begin{definition}[physically reasonable computer]
For any computer $C$,
we say that $C$ is physically reasonable if
there exist
$p,q\in\Dom C$
such that $\abs{p}\neq\abs{q}$.
\hfill\IEEEQED
\end{definition}

Then we can prove Theorem~\ref{inc-dec} below
in a similar manner to the proof of Theorem~7 of \cite{T09LFCS}.

\begin{theorem}\label{inc-dec}
Let $C$ be a physically reasonable computer.
Then each of the mapping $(0,1)\ni T\mapsto Z_C(T)$,
the mapping $(0,1)\ni T\mapsto E_C(T)$,
and the mapping $(0,1)\ni T\mapsto S_C(T)$ is
an increasing real function.
On the other hand,
the mapping $(0,1)\ni T\mapsto F_C(T)$ is
a decreasing real function.
\hfill\IEEEQED
\end{theorem}

In order to prove the main result,
it is also convenient to use the notion of \textit{computable measure machine},
which was introduced by Downey and Griffiths \cite{DG04} in 2004
for the purpose of characterizing the notion of Schnorr randomness of a real
in terms of program-size complexity.

\begin{definition}[computable measure machine, \cite{DG04}]
A computer $C$ is called a computable measure machine
if $\sum_{p\in\Dom C}2^{-\abs{p}}$ (i.e. $Z_C(1)$)
is computable.
\hfill\IEEEQED
\end{definition}

For the thermodynamic quantities in AIT,
we can prove Theorem~\ref{comp-prcmm} below
using Theorem~\ref{inc-dec}.

\begin{theorem}\label{comp-prcmm}
Let $C$ be a physically reasonable, computable measure machine.
Then, for every $T\in(0,1)$,
the following conditions are equivalent:
\begin{enumerate}
  \item $T$ is computable.
  \item One of $Z_C(T)$, $F_C(T)$, $E_C(T)$, and $S_C(T)$ is computable.
  \item All of $Z_C(T)$, $F_C(T)$, $E_C(T)$, and $S_C(T)$ are computable.\hfill\IEEEQED
\end{enumerate}
\end{theorem}

\begin{example}\label{ex-prc}
The following two computers are
examples of physically reasonable,
computable measure machines.

\noi\textbf{(i) Two level system:}
Let $B$ be a particular computer
for which $\Dom B=\{1,01\}$.
Then we see that, for every $T>0$,
\begin{align*}
  Z_B(T)&=2^{-1/T}+2^{-2/T},\\
  F_B(T)&=-T\log_2 Z_B(T),\\
  E_B(T)&=\frac{1}{Z_B(T)}\left(2^{-1/T}+2\cdot 2^{-2/T}\right),\\
  S_B(T)&=(E_B(T)-F_B(T))/T.
\end{align*}
\noi\textbf{(ii) One dimensional harmonic oscillator:}
Let $O$ be a particular computer for which
$\Dom O=\{0^l1\mid l\in\N\}$.
Then we see that, for every $T>0$,
\begin{align*}
  Z_O(T)&=\frac{1}{2^{1/T}-1},\\
  F_O(T)&=T\log_2 \left(2^{1/T}-1\right),\\
  E_O(T)&=\frac{2^{1/T}}{2^{1/T}-1},\\
  S_O(T)&=(E_O(T)-F_O(T))/T.
\end{align*}

Since $Z_B(1)=3/4$ and $Z_O(1)=1$,
we see that $B$ and $O$ are physically reasonable,
computable measure machines.
Note that Theorems~\ref{inc-dec} and \ref{comp-prcmm} certainly
hold for
each of
the particular
physically reasonable
computers
$B$ and $O$.
\hfill\IEEEQED
\end{example}

Based on Theorems~\ref{quc} and \ref{comp-prcmm},
the main result is proved as follows.

\begin{proof}[The proof of Theorem~\ref{main}]
We first choose
any optimal computer $V$ and
any physically reasonable,
computable measure machine $C$.
Then, for each $n\in\N^+$,
we denote the computer $V\oslash(C^{\oslash n})$ by $V_n$.
By Theorem~\ref{ocmpo},
we see that $V_n$ is optimal for every $n\in\N^+$.
Furthermore,
it is easy to see that
the infinite sequence $V_1,V_2,V_3,\dotsc$ of computers is recursive.
On the other hand,
it follows from Theorem~\ref{quc} that,
for every $n\in\N^+$ and every $T\in(0,1)$,
\begin{equation}\label{VnUOn}
\begin{split}
  Z_{V_n}(T)&=Z_{V}(T)Z_{C}(T)^n,\\
  F_{V_n}(T)&=F_{V}(T)+nF_{C}(T),\\
  E_{V_n}(T)&=E_{V}(T)+nE_{C}(T),\\
  S_{V_n}(T)&=S_{V}(T)+nS_{C}(T).
\end{split}
\end{equation}

Let $m$ and $n$ be arbitrary two positive integers with $m>n$.
Then it follows from the equations \eqref{VnUOn} that
\begin{align}
  Z_{V_m}(T)&=Z_{V_n}(T)Z_{C}(T)^{m-n},\label{ZVm-n}\\
  F_{V_m}(T)&=F_{V_n}(T)+(m-n)F_{C}(T),\label{FVm-n}\\
  E_{V_m}(T)&=E_{V_n}(T)+(m-n)E_{C}(T),\label{EVm-n}\\
  S_{V_m}(T)&=S_{V_n}(T)+(m-n)S_{C}(T)\label{SVm-n}
\end{align}
for every $T\in(0,1)$.
In what follows,
using \eqref{FVm-n}
we show that $\mathcal{F}(V_m)\cap\mathcal{F}(V_n)=\emptyset$.
In a similar manner,
using \eqref{ZVm-n}, \eqref{EVm-n}, and \eqref{SVm-n}
we can show that
$\mathcal{Z}(V_m)\cap\mathcal{Z}(V_n)
=\mathcal{E}(V_m)\cap\mathcal{E}(V_n)
=\mathcal{S}(V_m)\cap\mathcal{S}(V_n)
=\emptyset$
as well.

Now, let us assume contrarily that 
$\mathcal{F}(V_m)\cap\mathcal{F}(V_n)\neq\emptyset$.
Then there exists $T_0\in(0,1)$ such that
both $F_{V_m}(T_0)$ and $F_{V_n}(T_0)$ are computable.
It follows from \eqref{FVm-n} that
\begin{equation*}
  F_{C}(T_0)
  =\frac{1}{m-n}\left(F_{V_m}(T_0)-F_{V_n}(T_0)\right).
\end{equation*}
Thus, $F_{C}(T_0)$ is also computable.
Since $C$ is a physically reasonable, computable measure machine,
it follows from Theorem~\ref{comp-prcmm} that $T_0$ is also computable.
Therefore,
since $V_m$ is optimal,
it follows from Theorem~\ref{cprpffe} (i) that
$F_{V_m}(T_0)$ is weakly Chaitin $T_0$-random.
However,
this contradicts the fact that $F_{V_m}(T_0)$ is computable.
Thus we have $\mathcal{F}(V_m)\cap\mathcal{F}(V_n)=\emptyset$.
This completes the proof of Theorem~\ref{main}~(i).

Theorem~\ref{main}~(ii) and Theorem~\ref{main}~(iii) follow immediately
from Theorem~\ref{fptpr} and the fact that $V_i$ is optimal for all $i\in\N^+$.
\end{proof}




\section*{Acknowledgments}

This work was supported by
KAKENHI, Grant-in-Aid for Scientific Research (C) (20540134),
by SCOPE
of the Ministry of Internal Affairs and Communications of Japan,
and
by CREST of the Japan Science and Technology Agency.



%

\end{document}